\documentclass[a4paper,11pt]{amsart}
\usepackage{multicol,ifthen}
\usepackage{amssymb,stmaryrd}
\usepackage{harvard}
\usepackage{graphicx}

\begin{document}

\thispagestyle{empty}

\newcommand{\F}{\mathcal{F}}
\newcommand{\R}{\mathbb R}
\newcommand{\N}{\mathbb N}
\newcommand{\C}{\mathbb C}  
\newcommand{\h}[2]{\mbox{$ \widehat{H^{#1}_{#2}}$}}
\newcommand{\hh}[3]{\mbox{$ \widehat{H^{#1}_{#2, #3}}$}} 
\newcommand{\n}[2]{\mbox{$ \| #1\| _{ #2} $}} 
\newcommand{\x}{\mbox{$X^r_{s,b}$}} 
\newcommand{\xx}{\mbox{$X_{s,b}$}}
\newcommand{\X}[3]{\mbox{$X^{#1}_{#2,#3}$}} 
\newcommand{\XX}[2]{\mbox{$X_{#1,#2}$}}
\newcommand{\q}[2]{\mbox{$ {\| #1 \|}^2_{#2} $}}
\newcommand{\e}{\varepsilon}
\newcommand{\simle}{{^{\,<}_{\,\sim}}\,}
\newcommand{\simge}{{^{\,>}_{\,\sim}}\,}
\newtheorem{lem}{Lemma} 
\newtheorem{kor}{Corollary} 
\newtheorem{thm}{Theorem}
\newtheorem{prop}{Proposition}

\newcommand{\beq}{\begin{equation}}
\newcommand{\eeq}{\end{equation}}
\newcommand{\beqarr}{\begin{eqnarray}}
\newcommand{\eeqarr}{\end{eqnarray}}
\newcommand{\beqa}{\begin{eqnarray*}}
\newcommand{\eeqa}{\end{eqnarray*}}
\unitlength1cm

\hyphenation{Rie-man-nian}

\title{Scale covariant gravity and  equilibrium cosmologies }
\author[E. Scholz]{Erhard Scholz$\,^1$}\footnotetext[1]{\, scholz@math.uni-wuppertal.de\\
\hspace*{7mm}University Wuppertal, Department C, Mathematics and Natural Sciences,\\
\hspace*{6.5mm} and Interdiscplinary Center for Science and Technology Studies} 

\date{20. March 2007}

\begin{abstract}
Causal structure, inertial path structure and compatibility with quantum mechanics demand no full Lorentz metric, but only an integrable Weyl geometry for space time \cite{Ehlers/Pirani/Schild,Audretsch_ea}. A proposal of \cite{Tann:Diss,Drechsler/Tann} for a minimal coupling of the Hilbert-Einstein action   to   a scale covariant scalar vacuum field $\phi$ (weight $-1$) plus (among others) a
 Klein-Gordon action term opens the access to a scale covariant formulation of gravity.    The ensuing  scale covariant K-G equation specifies a natural scale gauge ({\em vacuum gauge}). Adding other  natural assumptions for gauge conditions (in particular Newton gauge, with unchanging 
Newton constant) the chosen Ansatz leads to a class of Weyl geometric
 Robertson-Walker solutions of the Einstein equation, satisfying $a''a+a'\,^2= const$, analogous to the Friedmann-Lemaitre equation but with completely different dynamical properties ($a$ the warp function in Riemann gauge).    The class  has an asymptotically attracting 1-parameter subfamily of  extremely  simple space-time geometries with an isotropic  Robertson-Walker fluid as source of the Einstein equation, discussed as {\em Weyl universes} elsewhere  \cite{Scholz:ModelBuilding}.
 Under the assumption of a  heuristic gravitational self energy binding Ansatz  for the  fluid,  equilibrium solutions arise, in stark contrast to classical (semi-Riemannian) cosmology.    Weyl universes agree very well  with a variety of empirical data from observational cosmology, in  particular  supernovae luminosities and quasar data.
\end{abstract}

\maketitle 

\section{Introduction}
Equilibrium models play no role in  present day cosmology. There are  strong reasons for this state of affairs in the classical framework of semi-Riemannian geometry, most importantly the  instability of classical static Robertson-Walker models, global singularity theorems, and the interpretation of cosmological redshift as a result of  space expansion. But these  are {\em no sufficient reasons} for an exclusive consideration of expanding space models. Already a tiny widening of the geometrical framework by integrable Weyl geometry     gives a completely different picture for mathematical  existence and physical acceptability of equilibrium cosmologies. They even stand in surprisingly good accord with a broad variety of empirical evidence of recent observational cosmology. Moreover, this widening of the geometrical framework   corresponds to a field theoretically natural extension of classical relativity and  cannot be pushed aside as a game of mathematical interest only. The following article explains why.

Although a general acquaintance with basic ideas of localized scale gauge is presupposed here, basic  notations and terminology of  Weyl geometry, which are  referred to in the sequel,  are introduced in    section 2. For the introduction of a scale gauge extension of classical relativity  we  follow  the work of Tann and Drechsler  with some adaptations (section 3). 
In particular  the observation of the authors that in Weyl geometry  a complex (or, more generally,  electroweak) scalar vacuum field $\phi$ offers excellent possibilities for forming scale invariant Lagrange densities  leads to a {\em scale covariant version of the Einstein equation}  \cite{Tann:Diss,Drechsler/Tann}. It  transfers directly to cosmology.

Variation of the scale invariant Lagrangian density with respect to $\phi$  (or its complex conjugate) gives a 
Klein-Gordon equation (section 4). Different to Drechsler/Tann we start from a 
Klein-Gordon action without external mass, usually regarded as a ``massless'' K-G field. Nevertheless there arise terms, which  look like mass terms but vary with space curvature. Conformal field theory knows this very well and  fixes specific  choices of coupling coefficients to ``get rid'' of these terms. In  Weyl geometry this is not necessary, due to its additional gauge freedom of scale. Here these terms express, in  a very natural way,  a    spontaneous acquirement of  mass by vacuum bosons through  coupling to gravity and a biquadratic
 self-interaction. This proposes a natural gauge for a Higgs mechanism coupled to gravity, which has been introduced  by  Drechsler and Tann in a slightly different setting (section 4).
Moreover, the spontanously acquired mass term of the vacuum field allows to  introduce a scale invariant mass parameter for any particle or quantum by comparison with the vacuum boson mass. It  leads to a physically  very plausible gauge ({\em vacuum gauge}) and gives reason to reconsider, at least  in the cosmological context, Weyl's old gauge hypothesis (section 5). 

Next we consider Robertson-Walker geometries in this framework (section 6). The gauge conditions from section 4 and a condition satisfied by the main example of condition of section 7 specify a  class of Weyl geometric Robertson-Walker geometries, which   fulfill $a''a+a'\,^2= const$  ($a$ the warp function in Riemann gauge). This class has  dynamical properties which are completely different from Friedmann-Lemaitre models.    In particular, there exists  an asymptotically attracting 1-parameter subfamily of  intriguingly simple space-time geometries, discussed as {\em Weyl universes} elsewhere, e.g., \cite{Scholz:ModelBuilding}. 
The latter are derived from the well known ``static'' Robertson-Walker manifolds by superimposing a constant scale connection which expresses cosmological redshift ({\em Hubble connection}).

We now turn to the r.h.s. of the Einstein equation and study properties of highly symmetric 
 scale covariant isentropic fluids. It is possible to  find a mathematically consistent expression for negative self binding energy, dependent on mass density (``case 2'', section 7). That leads to a link between the vacuum energy tensor and mass energy density, which may be of wider import than just this Ansatz  (section 8). It ensures a {\em stable equilibrium solution} for the corresponding Robertson-Walker fluid solution of the  Einstein equation (section 9).

Such equilibrium solutions are of  cosmological interest only, if they are considered in the Weyl geometric framework. Then they go together with cosmological redshift. The material interpretation of the case 2 fluid  of section 7 leads to a satisfying implementation of the Mach principle. The inertial and metrical structure of Minkowski space arises here by abstraction from a Weyl universe with critical mass density (section 10). 
 The article ends with  short remarks about empirical aspects of {\em Einstein-Weyl}  universes (Weyl universes of positive sectional curvature). It concludes that, in the light   of the Weyl geometric perspective, equilibrium solutions of cosmology should neither  be neglected for cosmological theory building, nor  for the analysis  and interpretation of data from observational cosmology (section 10).

\section{Geometric preliminaries}
Weyl geometry  has a  long tradition. It  started  with H. Weyl's  extension of (semi-)Riemannian geometry by a localized scale gauge 
 \cite{Weyl:InfGeo,Weyl:Erweiterung}. By many reasons, most importantly coherence with quantum physics  \cite{Audretsch_ea},   the restricted version of an  integrable scale connection ({\em integrable Weyl geometry}, IWG) is now exclusively used for physical purposes.  P.A.M. Dirac enhanced it conceptually in an early attempt to bridge the gap between nuclear forces (the strong force) and gravity \cite{Dirac:1973}. Once in a while and in varying contexts, IWG has been taken up by scientists pursuing different research programs in mathematical physics or geometry.  To list just a few:\footnote{There
 are many more works on Weyl geometry in mathematical physics and geometry; the list raises no claim for representativity or even completeness.  By well known reasons, even Weyl's own early intended application of WG as a unifying framework of gravity and electromagnetism has been omitted from this short list.}
 \cite{Ehlers/Pirani/Schild} based their  foundational and conceptual clarification of general relativity upon it. \cite{Canuto_ea} and \cite{Maeder:WeylGeometry} followed the line of  field theoretic and cosmological investigations opened up by Dirac. This program produced interesting partial results   and  has found  active protagonists until the present  \cite{Tiwari:Weyl_Geometry,Israelit:Weyl_DiracGeometry}.  \cite{Tiwari:Electron} tries to understand electron matter by Weyl geometry.
 \cite {Santamato:WeylSpaces,Santamato:KG}, on the other hand, started a new approach which  links the length scale with  Bohm-type guiding fields and quantum potentials. That idea has been taken up  by 
\cite{Castro:WeylGeometry,Castro:WeylGeometry2006} and was extended to questions of geometric quantization. In  \cite{Castro:DarkEnergy} even a mixed 
Brans-Dicke and Weyl geometric approach has been used in search for understanding ``dark energy''.  \cite{Tann:Diss}  and \cite{Drechsler/Tann,Drechsler:Higgs} have proposed a semiclassical  field theoretic approach to understand  mass generation by the breaking of Weyl symmetry. Finally
\cite{Folland:WeylMfs,Varadarajan:Connections} and others  discuss Weyl geometry  from the point of view  of  differential geometry. Studies in conformal geometry like \cite{Frauendiener:Conformal} are helpful for technical questions of rescaling. 

Already this sporadic list shows that the conceptual extension of Riemannian geometry, made possible by Weyl's local scaling idea, has no canonicaöl and unique physical application. Its intriguing mathematical and conceptual design can be  made fruitful for a variety of purposes, with different chances for success. Here our main goal is to  gain a broader, and perhaps deeper, understanding of cosmological redshift and its relation with the vacuum structure, similar to a proposal made public by P. Cartier in his talk \cite{Cartier:Cosmology}. From the variety of works quoted, Tann's and Drechsler's contain ideas and results which link most closely to our perspective.

A general  acquaintance with Weyl geometry will be be presupposed.\footnote{See, among others,  \cite{Weyl:Erweiterung,Bergmann:Relativity,Dirac:1973,Canuto_ea,%
Israelit:Weyl_DiracGeometry,%
Eisenhart:1926,Folland:WeylMfs,Varadarajan:Connections,%
Scholz:Extended_Frame,%
Scholz:ModelBuilding}. {\em Erratum:} In formulas (5), (6) for Ricci and scalar curvature of \cite{Scholz:ModelBuilding} the covariant derivative $\nabla $ has to be taken with respect to the Riemannian component of the gauge $(g,\varphi)$ only, $\nabla = \; _g\hspace{-3pt}\nabla $.}
 Here only the   main terminological conventions and  notations used  in the 
sequel can be listed.

We work on  differentiable manifolds $M, \; \dim M = 4$, endowed with a Weylian metric, i.e., an equivalence class $[(g,\varphi)]$ of pairs $(g, \varphi)$ constituted by a 
semi-Riemannian metric $g$ of Lorentz signatur $(-,+,+,+,)$ and a differential 1-form $\varphi$. In local coordinates the latter are given by $g_{\mu \nu }$ and $\varphi_\mu $. A choice of $(g,\varphi)$ is called a  (scale) {\em gauge} of the metric. A change of representative from $(g, \varphi)$ to $(\tilde{g}, \tilde{\varphi})$ given by
\[  \tilde{g} =  \Omega ^2 g \; , \quad \quad  \tilde{\varphi} = \varphi - d \log \Omega \] 
is called  {\em (scale)  gauge transformation}, where $\Omega > 0$ is a strictly positive real function on $M$.  $g$ is  the {\em Riemannian component} of the Weyl metric and $\varphi$ its {\em scale} (or length) {\em  connection}. The connection is {\em integrable}, iff 
\[  d \varphi = 0 \; .\]
By reasons indicated above, we only work  with {\em integrable Weyl manifolds}. 
For simply connected  $M$ there exists a gauge in which the length connection vanishes, $\tilde{g} = \lambda ^2 g$,  $\tilde{\varphi}= 0$, with $\lambda (x) = e^{\int_{x_0}^x  \varphi(c'(u))du}$, $c(u)$ any differentiable  path  from a fixed reference point $x_0$ to $x$. This gauge is called {\em Riemann gauge} (in the physical literature often also called {\em Einstein gauge}), and $\lambda$ the {\em length transfer function} of the gauge $(g,\varphi)$. That allows to look at the integrable Weyl manifold from the point of view of (semi-)Riemannian geometry, but it does not force us to do so. 

The concepts of differential  geometry, known from the (semi-) Riemannian case, can be transferred to the Weyl geometric case. The calculation of  the 
{\em covariant derivative} $ D_{([(g,\varphi)]}\, F$  of an ordinary, i.e., scale invariant, vector or tensor field $F$ with respect to the Weylian metric $[(g,\varphi)]$ includes terms in $\varphi$, but the result is a scale invariant vector or  tensor field.  The same is true for Weylian {\em geodesics} $\gamma_W$ and for the  {\em curvature tensor} $R = R^{\alpha} _{\beta \gamma \delta}$,  which turn out to be invariant under scale gauge transformations.
But often   we need concepts which allow to compare metrical measurements at different points of $M$ in different gauges. For this purpose the length transfer function $\lambda $ can be used.\footnote{More details in \cite{Scholz:Extended_Frame}.}

The Weyl structure of $M$ allows to consider (real, complex etc.) {\em Weyl functions} $f$  and (vector, tensor, spinor \ldots) {\em Weyl fields} $F$ on $M$, which transform under gauge transformations by
\[  f \longmapsto \tilde{f} = \Omega ^k f \;  ,\quad \quad F \longmapsto \tilde{F} = \Omega ^l F  \; . \]
$k$ and  $l$ are the (scale or Weyl) {\em weights} of $f$ respectively $F$.\footnote{Mathematically, Weyl functions and Weyl fields are themselves equivalence classes of ``ordinary'' (scale invariant) functions and fields. }
We also write 
\[  [[f]]:= k \; , \quad \quad [[F]] := l  \]
for the Weyl weight of functions or fields. With the curvature tensor $R = R^{\alpha} _{\beta \gamma \delta}$  of the Weylian metric also  the Ricci curvature tensor $Ric$ is scale invariant, while  scalar curvature 
\[  \overline{R} = g^{\alpha \beta } Ric\, _{\alpha \beta } \]
is of weight $ [[ \overline{R} ]] =-2$.

For any nowhere vanishing Weyl function $f$ on $M$ with weight $k$ there is a gauge (unique up to a constant), in which $\tilde{f}$ is constant. It is given by 
\beq \Omega = f^{-\frac{1}{k}} \;  \label{f gauge}
\eeq
 and will be called   {\em f-gauge} of the Weylian metric. There are infinitely many gauges; some of them are of particular importance. An 
$\overline{R}$-gauge (in which scalar curvature is scaled to a constant) exists for manifolds with nowhere vanishing scalar curvature. It will be called   {\em Weyl gauge}, because Weyl assigned it a particularly important role in his foundational thoughts about  matter and  geometry  \cite[298f.]{Weyl:RZM5} (cf. section 4). 

Any semi-Riemannian manifold can be considered in the extended framework of (integrable) Weyl geometry. Contrary to a widespread opinion it  makes sense, under certain circumstances, to do so. The rest of this article will give a first taste. For  cosmological studies, Robertson-Walker manifolds are particularly important. There we have the diffeomorphism
\[  M \approx  R \times S_{\kappa } \]
with $S_{\kappa }$ a  Riemannian space of constant sectional curvature $\kappa $, here usually (but not necessarily) simply connected. If in spherical coordinates $(r,\Theta, \Phi )$ on $S_{\kappa }$
\beq d \sigma _{\kappa } ^2  =  \frac{dr^2}{1- \kappa r^2} +r^2 (d\Theta ^2  + \sin^2 \Theta \, d \Phi^2) \label{metric constant curvature} \eeq 
 denotes the metric on  the spacelike fibre, the Weylian metric $[(g, \varphi )]$ on $M$ is specified by its Riemann gauge $(\tilde{g},0)$ like in standard cosmology:  
\beq \tilde{g}: \quad \quad ds^2 = - d{\tau }^2 + a(\tau )^2 d \sigma _{\kappa } ^2 \; \label{Robertson Walker metric}\eeq 
Here $\tau = x_0 $ denotes  a well chosen local or global coordinate (cosmological time parameter of the semi-Riemannian gauge) in $\R$, the first factor  of $M$. In the semi-Riemannian perspective 
$a(\tau )$, the {\em warp function} of $(M, [(g,\varphi)])$, is usually interpreted as an expansion of space sections.  The Weyl geometric perspective shows that it need not. For example, for any Weyl-Robertson-Walker manifold there is a gauge$(g_w, \varphi_w)$   in which the ``expansion  is  scaled away'' by 
\beq 
g_w = \Omega _w^2 g \quad \mbox{with } \;\;  \Omega _w := \frac{1}{a} \; .
\eeq
With 
\[  t:= \int ^{\tau } \frac{du}{a(u)} = h^{-1}(\tau ) \quad \mbox{and its inverse function} \;\; h(t) = \tau \] we   get a ``static'' metric
\beq g_w(x): = -dt^2 + d\sigma _{\kappa }^2 \;   \quad\;\;  \varphi_w (x) =  - d \log (a \circ h)  = (a'\circ h) \, dt = a'(\tau (t)) dt \; .   \label{warp gauge} \eeq  
 It will be called the {\em warp gauge} of the Robertson-Walker manifold.\footnote{In earlier publications also called ``Hubble gauge'',   \cite{Scholz:ModelBuilding}.} 
Here the cosmological redshift is no longer mathematically characterized by a warp function $a(x_0)$  but by the scale connection $\varphi_w$. We shall call $\varphi = \varphi_w$ the {\em Hubble connection} of the Weyl manifold (in warp gauge) \cite{Scholz:ModelBuilding}.

Dirac extended the calculus on integrable Weylian manifolds by defining scale covariant derivatives  of Weyl functions or Weyl fields $F$ of any weight $[[F]]$
\[ D\, F:=  D_{[(g,\varphi)]} F + [[F]] \,  \varphi \otimes F \; . \]
They are Weyl tensor fields (one order higher than $F$) of unchanged scale weight $[[D \, F]] = [[F]]$.  For the description of relativistic trajectories Dirac introduced {\em scale covariant} geodesics $\gamma _D$. They arise from Weyl's scale invariant geodesics $\gamma _W$ by reparametrization, such that the weight of the tangent field $u:= \gamma_D '$ is $[[u]]-1$. Then  $g(u,u)$ is scale invariant,  and with it the  distance measurement by Diracian geodesics, which coincides with Riemannian distance. In this sense the danger arises that, in the Diracian approach, the gauge aspect is not taken sufficiently serious from a metrical point of view.\footnote{Don't identify physical metric with the gauge invariant distances!}
 But it offers the great advantage that  Dirac's scale covariant geodesics have the same scale weight as energy $E$ and mass $m$, $[[E]] = [[m]]= -1$. Therefore mass or energy factors assigned to particles or field quanta can be described more easily in a gauge independent manner in Dirac's calculus. This was Dirac's important contribution for facilitating  the acceptance of other gauges than Riemann gauge for physical purposes, in particular for a scale covariant theory of gravity.

\section{Scale covariant gravity}
A {\em causal structure} of space-time $M$, $\dim M = 4$, is specified by a conformal structure  $[g]$ of  a Lorentz type  semi-Riemannian metric $g$ on $M$ ($sign\, (g)=(3,1)$). An {\em inertial structure} on $M$ is given by a projective path structure $\{ [ \gamma ] \}$ of (arbitrarily parametrized) paths $\gamma $ with any initial condition in $TM$,  the tangent bundle of $M$. If the causal structure and the inertial structure satisfy certain compatibility 
conditions,\footnote{Compatibility conditions are analyzed more closely in \cite{Ehlers/Pirani/Schild}. Most importantly: If the initial condition of a path $\gamma $ lies ``on''  the boundary $\partial \mathcal{C} $ of a cone $\mathcal{C}$ of $[g]$, the whole path $\gamma $ is in $\partial \mathcal{C}$.} $[g]$ and $\{ [ \gamma ] \}$  specify uniquely  a {\em Weylian metric} $[(g,\varphi )]$ on $M$ \cite{Weyl:Projektiv_konform,Ehlers/Pirani/Schild}. Compatibility of the gauge metric structure  $[(g,\varphi )]$ with quantum mechanics demands  the constraint of  {\em integrability } of the metric, $d \varphi = 0$, \cite{Audretsch_ea}.\footnote{Compatibility of QM with $ [(g, \varphi)])$  is specified by the condition that the WKB development of a massive Klein-Gordon field on $(M, [(g, \varphi)])$ coincides with geodesics of the metric structure, up to  first order.}
Although the  Weylian metric (integrable or not) is a fullfledged metrical structure in the mathematical sense, its scaling freedom signals an underdetermination from the physical point of view. 
Obviously it is necessary to  specify  a preferred scale gauge $(g_o, \varphi_o)$ for observable quantities, in order to arrive at a metrically well defined physical geometry. It will be called the {\em observational gauge} of the 
theory.\footnote{In \cite{Scholz:ModelBuilding} it has been called ``matter gauge'' because of its  directly visible effects on material measurements.}

Different proposals have been presented  for the choice of observational gauge in the literature. Usually they result in  Riemann gauge.\footnote{\cite{Ehlers/Pirani/Schild,Audretsch_ea,Drechsler/Tann} }
But Dirac's  scale covariant geodesics, combined with scale covariant derivation, show that this is not imperative (see end of last section).  Here we take up the proposal of scale fixing by a scale covariant (complex) scalar vacuum Weyl field $\phi$  on $M$ with weight $[[\phi ]] = -1$ \cite{Drechsler/Tann} in  a slightly modified form.   A mass term of $\phi$ acquired spontaneously by its specific coupling to gravity will give a natural specification for observational gauge  (``vacuum gauge'', see below). The Lagrangian is chosen such that $\phi$ couples to the Einstein-Hilbert metric in the most simple way to achieve a {\em scale  invariant} Lagrangian density. In this way, the vacuum field is related to the metric by an adapted kind of  ``minimal''  coupling. Both together, the metric $[(g, \varphi)]$ and the vacuum field $\phi$, specify a scale covariant theory of gravity including a full metrical determinaton of the metric (i.e., specification of observational gauge).

In this sense we study the dynamics on a differentiable manifold $M$ endowed with  Weylian metric $[(g, \varphi )]$, governed by  a  Lagrangian  density $\mathcal L$ and action
\beq {\mathcal S} = \int  {\mathcal L} dx = \int L(g, \partial g, \varphi, \phi , \phi ^{\ast}, F, \rho ) d \omega _g    \;  .  \label{total action}  \eeq
    $F = (F_{\mu \nu })$ is a scale invariant $(2,0)$ tensor field (electromagnetic field), $\rho$ a real field of weight $-4$ (matter density), $\phi $   a complex scalar field (vacuum field)  and $\phi ^{\ast}$ its complex conjugate, both of weight $[[\phi]] = -1$. 
We use the notation
 \[ d \omega _g   =  \sqrt{|g|} dx \, , \quad |g| := |det\; g| \,,  \]
 for the volume form of the Riemannian component $g$ of the metric.  
In order for the Lagrangian  $\mathcal L$  to be  scale invariant, $L$ has to be of 
 Weyl weight $[[L]]=-4$, because $|g|$ transforms under rescaling by weight  $+8$.
  
Following \cite{Tann:Diss} and \cite{Drechsler/Tann}, we  work with a Lagrangian density ${\mathcal L} = L  \sqrt{|g|}$, which   contains  the following scale covariant versions of well known   Lagrangians:
\beqa  L_{HE} &=& \alpha \, \overline{R} \, (\phi ^{\ast} \phi) \; \;  \quad \quad \quad \mbox{Hilbert-Einstein action } \\
L_{\Lambda } &=& - 2\, \alpha \beta \, (\phi ^{\ast} \phi) ^2 \quad  \quad \mbox{cosmological term (factor $-2 \alpha $ for convenience) } \\
L_{KG} &=& \gamma  \, D^{\mu }\phi^{\ast}D_{\mu} \phi \quad \;\;\; \;\;\; \mbox{scalar field action without (external) mass} \\
L_{em} &=& \delta \,  F_{\mu \nu } F^{\mu \nu } \;  \quad \quad \quad \;\; \mbox {electromagnetic action } \\
L_m &=& L_m (\rho ) \; \; \;\; \quad \quad \quad \quad \mbox{matter action } 
\eeqa
 $\alpha , \beta , \gamma , \delta $  are scale invariant coupling constants; therefore   all contributions of $L$ are of  scale weight $-4$.  $L_m $ is a {\em  classical action} of a cosmic matter flow; it depends exclusively on matter density $\rho $ (not on $\phi$).

Our Lagrangian differs crucially from   Weyl's original Ansatz which used a quadratic term in $\overline{R}$ to obtain the correct weight. In our approach the Hilbert-Einstein term appears coupled to the vacuum field. That leads  to  scale invariance without raising the power of $\overline{R}$.

The total Lagrangian is: 
\beqarr
{\mathcal L} &=& {\mathcal L}_{HE} +  {\mathcal L}_{\Lambda } +    + {\mathcal L}_{KG} +  {\mathcal L}_{em}  +  {\mathcal L}_m  = L   \sqrt{|g|} \; , \label{total Lagrangian} \\
 L &=&  \alpha \, \overline{R} (\phi ^{\ast} \phi)   - 2 \alpha \beta \, (\phi ^{\ast} \phi) ^2 +   \gamma  \, D^{\mu }\phi^{\ast}D_{\mu} \phi + \delta  \,  F_{\mu \nu } F^{\mu \nu }    + L_m  \; ,  \nonumber
\eeqarr
where $\frac{\delta {\mathcal L}_m}{\delta \phi } = 0$.
It  has common  features  with the theories of  Dirac and of  Brans-Dicke, but differs from both. Dirac introduced a ``Lagrangian multiplier'' $\beta $ of scale weight $[[\beta ]]= -2$, which formally played the role of $\phi^{\ast} \phi $. He used it for a peculiar search for gauges expressing his ``large number hypothesis''. 
Brans-Dicke's  scalar field action contains  an additional factor of type $  (\phi^{\ast} \phi) ^{-1}$.  Similar to Tann's and Drechsler's, our scalar field action is that of a 
 Klein-Gordon field,  $ {\mathcal L}_{KG} $, but here without external mass. That is an important difference to Brans-Dicke theory. In addition, wie have reasons for the estimation $\gamma  \ll  ||\rho _{crit}||$(section 4),  in strong contrast to Brans-Dicke theory (where $\gamma \sim 1$). 

On the other hand,   the Lagrangian has a close kinship to conformal Klein-Gordon field theory  for the massless case, ${\mathcal L}_m= 0$.  For $\gamma = \frac{2(n-1)}{n-2} \, \alpha , \; n = \dim M $,  it even {\em is} conformally invariant 
\cite[equ. (190)]{Tann:Diss},  \cite[394f.]{Carroll:Spacetime}, $\gamma = 3 \, \alpha $ for $n=4$.  Tann and Drechsler have taken this as the starting point, for what they call ``mass generation by breaking Weyl symmetry'' \cite{Drechsler/Tann}. That is an interesting point of view; but their research program makes it necessary to introduce an ad-hoc external mass term into the Lagrangian of the Klein-Gordon field (see below).   
Here we start more modestly, also more in line with Weyl geometry, without any prior specification of  $\gamma $. 
The conformal invariance is broken  already for ${\mathcal L}_m= 0$ and the 
 invariance group of the total Lagrangian (including mass term) remains the (localized) scale extended Lorentz group $G= \R^+ \times SO(1,3)$,  i.e., the  gauge group of a principal bundle over $M$ with fibre $G$.  If things go  well,  $\gamma $ may be   be specified at a  later stage. In the next section we shall see  that   the vacuum field $\phi$  specifies a naturally distinguished gauge. This has nothing to do with ``symmetry breaking''; it rather  allows a {\em natural fixing of the gauge}.

Weyl geometric terms in $\varphi$  come into the play only for the variation of ${\mathcal L}_{HE}$  and  ${\mathcal L}_{KG}$ with respect to $g$. The variation for the (semi-) Riemannian case   is well known from standard literature on GRT, e.g. \cite{Carroll:Spacetime,Straumann:Relativity,Weinberg:Cosmology,Hawking/Ellis}. Calculations  of the Weyl geometric case in \cite{Tann:Diss,Drechsler/Tann}  show that even for $L_{HE}$ and $L_{KG}$  the results are similar  in form to those in Riemannian geometry, although now they contain terms in the scale connection.   After adaptation of coefficients and if necessary  signs  because of signature change, we arrive at:
\beqa 
\frac{\quad \delta {\mathcal L}_{HE}}{  \sqrt{|g|} \delta g^{\mu \nu }} &=& \alpha  \phi ^{\ast} \phi \, (Ric - \frac{\overline{R}}{2} g)_{\mu \nu } \quad \quad  \quad \quad \quad \mbox{\cite[(2.16)]{Drechsler/Tann}}  \\
 \frac{\quad \delta {\mathcal L}_{\Lambda }}{\sqrt{|g|} \delta g^{\mu \nu }} &=&\alpha  \beta \, (\phi ^{\ast} \phi )^2 g_{\mu \nu }  \quad \quad  \quad \quad \quad  \quad \quad  \quad   \mbox{\cite[(2.16)]{Drechsler/Tann}} \\
 \frac{ \quad \delta {\mathcal L}_{KG}}{  \sqrt{|g|} \delta g^{\mu \nu }} &=& \gamma \left( D_{(\mu } \phi ^{\ast} D_{\nu )} \phi  -D_{(\mu } D_{\nu )}(\phi ^{\ast} \phi ) - g_{\mu \nu } \left( D^{\lambda } D_{\lambda } (\phi ^{\ast} \phi ) - \frac{1}{2} D^{\lambda } \phi ^{\ast} D_{\lambda } \phi   \right) \right) \\
\quad  & & \quad    \mbox{\hspace*{47mm}  \cite[(2.17)]{Drechsler/Tann} } \\
 \frac{\quad \delta {\mathcal L}_{em}}{\sqrt{|g|}  \delta g^{\mu \nu }} &=& 2 \delta  \, \left( - F_{\mu \lambda } F^{\lambda}_{ \nu } +\frac{1}{4} g_{\mu \nu  } F_{\kappa \lambda }  F^{\kappa \lambda }\right)\\ & & \mbox{\hspace*{51mm}  \cite [(2.18)]{Drechsler/Tann} } 
\eeqa
(Compare  \cite[(2.90), (2.101)]{Straumann:Relativity}, or others.) 
In order to connect to established knowledge on couplings, we demand that
\beq  \alpha = (8 \pi g_N)^{-1} [c^4] \, , \quad \delta = (8 \pi )^{-1} \label{alpha substitution}
 \eeq
($g_N$ Newton constant; in the sequel we suppress factors in the velocity of light, $c$). We thus arrive at the following {\em gauge covariant form of the Einstein equation}
\beq Ric - \frac{1}{2} \overline{R} g =  8 \pi \, g_N ( \phi ^{\ast} \phi )^{-1} \left( T^{(m)} + T ^{(em)} + T^{(KG)} \right) + T^{(\Lambda )}     \; \label{Einstein equation}
\eeq
with r.h.s terms:
\beqarr 
T^{(m)}_{\mu \nu } &=& - \frac{1}{\sqrt{|g|} } \frac{\delta {\mathcal L}_{m}}{\delta g^{\mu \nu }}  \quad \quad \quad \quad \quad \quad \quad \quad \quad \;\;  \; \mbox{matter tensor } \label{general matter tensor}\\
T_{\mu \nu } ^{(em)} &=& \frac{1}{4 \pi}  \left(  F_{\mu \lambda } F^{\lambda}_{ \nu } -\frac{1}{4} g_{\mu \nu  } F_{\kappa \lambda }  F^{\kappa \lambda }\right) \quad \quad \; \;\mbox{e.m. energy stress } \\
T^{(KG)} _{\mu \nu } &=& \gamma \bigg(  D_{(\mu } D_{\nu )}(\phi ^{\ast} \phi ) - D_{(\mu } \phi ^{\ast} D_{\nu )} \phi 
 + g_{\mu \nu } \Big( D^{\lambda } D_{\lambda } (\phi ^{\ast} \phi )     
 \\ & &  \quad \quad \quad  \quad  \quad - \frac{1}{2} D^{\lambda } \phi ^{\ast} D_{\lambda } \phi   \Big) \bigg)  
 \quad \quad \quad \mbox{ K-G energy stress } \label{KG tensor} \nonumber \\
T^{(\Lambda )} _{\mu \nu } &=& - \beta   (\phi ^{\ast} \phi ) \, g_{\mu \nu } \quad  (= - \Lambda  \, g_{\mu \nu } ) \quad \quad \quad \quad  \mbox{ $\Lambda $-tensor (cf. (\ref{Lambda and beta}))} \label{vacuum tensor}
\eeqarr
The left hand side (l.h.s.) of equation (\ref{Einstein equation}) is gauge
 {\em invariant}. The building blocks of the r.h.s are  gauge covariant only, but  the gauge weights of the factors cancel,  $[[\phi ^{\ast} \phi )^{-1}]] = 2, [[T ]] = -2 $. Thus the whole r.h.s. is  gauge invariant, as it must be. 
\beq  G_N:= g_N ( \phi ^{\ast} \phi )^{-1} 
\eeq 
may be considered  as {\em scaled version of Newton's gravitational constant}. It is a scalar Weyl function of weight 2 in accordance with the dimensional weight of $g_N$. In {\em this respect } our approach takes up a common motif of Dirac and of 
Brans-Dicke theory \cite[929]{Brans/Dicke} (also of \cite{Scholz:ModelBuilding}). Don't forget, however, that neither the form of the scalar field action nor the order of magnitude of the coupling coefficient   coincides with the Brans-Dicke approach.  As a consequence, the modification of classical general relativity is here much less drastic than in the latter. 

A cosmological specification of our approach in terms of Weyl universes (cf. section  \ref{section cosmological mean geometry}) has low velocity and weak field approximations which, expressed by parameters of parametrized post-Newtonian gravity \cite{Will:LivingReviews}, leads   to    $\alpha _1, \ldots , \alpha _4  \approx   0$, $  \zeta _1 , \ldots , \zeta _3 \approx   0, \; \beta , \gamma \approx  1$, up to terms at the order of cosmological magnitude ($\sim  H_0$) \cite[16f]{Scholz:ModelBuilding}. Thus dynamical predictions  agree,  inside the observational error margins, with those of classical relativity  and with the results of high precision observations. An exception may be seen, at first sight, in the anomalous frequency shift of the Pioneer spacecrafts. It finds an extremely easy explanation in the Weyl geometric framework \cite{Scholz:Pioneer}. But  it  turns out  to be of non-dynamic origin and is therefore no exception to the statement  on basic dynamical agreement of Weyl universe dynamics with classical relativity  on the solar system level. 

With
\beq  \Lambda :=  \beta \,  (\phi ^{\ast} \phi ) \label{Lambda and beta}
\eeq
we get the known form 
\[ T^{(\Lambda )} =  - \Lambda \, g \] 
for the  r.h.s. ``cosmological'' term.
Here $\Lambda $ is no ``true'', i.e., gauge invariant constant, but a scalar Weyl function of weight -2 (it has to be, in order to arrive at a gauge invariant term $\Lambda \, g $). The dependence of $\Lambda $ on the norm of the gravitational vacuum field $\phi$ expresses an indirect relationship between the $\Lambda $-term and the 
mass-energy and field energy content of the r.h.s. of (\ref{Einstein equation}). Below we shall see that  (in vacuum gauge) $\phi$ varies with the scalar curvature of the metric.  At least under specific conditions (case 2, section \ref{section mass term}), $T^{(\Lambda )}$  may be considered as  an expression for  self-energy and stresses of the gravitational field. It would be interesting to see, whether this interpretation can be generalized.

 Up to a (true) constant there is exactly one gauge in which the norm of the  vacuum field is constant. It is often helpful, although not always necessary, to normalize  such  that 
\beq \phi ^{\ast} \phi  = const = 1 \; . \label{Newton gauge}
\eeq 
By obvious reasons  this choice,  canonically specified by the  vacuum field,  will be called the {\em Newtonian constant gauge} or simply {\em Newton gauge}. In this gauge we have
\[ G_N = const = g_N \; . \]
Contrary to a widespread belief, Newton gauge {\em need   not} necessarily coincide with Riemann gauge of the Weylian metric. 

In any application of Weyl geometry it remains the question which gauge should be considered as the one which expresses empirical measurements and their scales. At the beginning of this section it has been introduced as observational gauge. Standard gravity has considered it self-evident that Riemann gauge {\em is} the observational gauge. Combining this identification with the empirically well corroborated observation that the Newton constant does not change, even over cosmological time  \cite{Will:LivingReviews} and spatial distances,  a silent claim of the standard approach becomes apparent. 
It can be stated as the following identification:\\[1.5mm]
{\bf Silent claim of (semi-)Riemannian gravity (SCRG).}  {\em Observational gauge coincides with 
 Riemann gauge  and  Newton gauge (constant $G_N$). In Weyl geometric terms }
\beq \varphi = 0    \Longleftrightarrow \phi ^{\ast} \phi  = 1 \label{central dogma} \eeq

In fact, this is nothing but a hidden hypothesis. We shall see that there are good reasons to admit alternatives and to study them seriously.
\vspace{2mm}

From the semi-Riemannian case we know  that in Riemann gauge
\[  div \; g^{-1} (Ric -\frac{\overline{R}}{2}g )  = 0 \; \]
in the sense of covariant divergence $div$. 
Scale-covariant differentiation of scale covariant tensors leads to scale covariant tensors, and the Einstein tensor is even scale invariant. Thus the result holds for any gauge and we find infinitesimal energy  momentum conservation in the sense of
\beq div \; T = 0 \; ,
\eeq 
with $T$ the total r.h.s energy momentum tensor of (\ref{Einstein equation}). 

 $\varphi$ is no dynamical field in the integrable Weyl geometric  frame,  but a 
non-dynamic  gauge  freedom for the metric only. Therefore an independent variation of $\varphi$ is not meaningful. Its  observationally distinguished value  is  fixed by a  gauge condition imparted by  the vacuum field $\phi$  which we turn to next.

\section{The vacuum field}
Variation with respect to $\phi$ or  with respect to $\phi^{\ast}$, is much more illuminating. 
 H. Tann and W. Drechsler have alrady discussed the variation of (\ref{total action}) with respect to $\phi^{\ast}$. In the absence of strong quantum fields (outside of matter concentrations) we have
\[  \frac{\delta {\mathcal L}_m}{\delta \phi^{\ast}} =0 \;  \]
and arrive at the following scale covariant  equation for the vacuum field \cite[equ. (358)]{Tann:Diss}, \cite[(2.13)]{Drechsler/Tann}:\footnote{Drechsler and Tann do not consider the massless case but add an ad hoc external mass term; see below.}
\beq D^{\mu} D_{\mu } \phi = \gamma ^{-1} \alpha  \left( \frac{\overline{R}}{2} - 2 \beta  (\phi ^{\ast} \phi ) \right) \phi
\eeq
Of course the situation becomes more involved if we study regions inside stars or even  collapsing objects. Then a coupling to quantum matter fields may become indispensable.

The r.h.s factor \footnote{In the presence of quantum matter there may be additional terms.} 
\beq M_0^2 :=   \gamma ^{-1} \alpha \left(  \frac{\overline{R}}{2} - 2 \beta  (\phi ^{\ast} \phi ) \right) =  \gamma ^{-1} (8 \pi g_N)^{-1}\left(  \frac{\overline{R}}{2} - 2 \Lambda  \right) \label{Klein-Gordon mass}
\eeq 
is equivalent to a quadratic mass term of a Klein-Gordon equation with  mass $m_0$, where $M_0 = m_0 c \, \hbar^{-1}$ (in inverse length units). It is 
 {\em   acquired  spontaneously} by the vacuum field, due to its coupling to gravity, $\alpha \overline{R} (\phi ^{\ast} \phi ) $, and its biquadratic self interaction  $-2\alpha  \beta (\phi ^{\ast} \phi )^2 $.

Its scaling behaviour is what we expect from mass,  $[[M_0 ]]=[[m_0]]= -1$,  respectively energy $E$, if we construct our theory such that the Planck constant is a true constant (which implies $[[E]]=-[[T]]$ because of $E = \hbar \nu $). Therefore in Weyl geometry  there is no reason to ``define it away'' like in the conformal approach.  The scale dependent mass terms even specify a distinguished gauge in which $M_0$ becomes constant (see below, vacuum gauge).

Although the curvature terms in the brackets of (\ref{Klein-Gordon mass}) are cosmologically small, $m_0$ may be  considerable or even large, depending on the order of magnitude of the inverse coupling factor $\gamma ^{-1}$. From observational cosmology  we have learned  that the term ${\mathcal L}_{KG}$ is negligible for cosmological calculations. All we need to arrive at observationally valid cosmological models is ${\mathcal L}_{HE}+{\mathcal L}_{m}+{\mathcal L}_{\Lambda }$. 
Drechsler's and Tann's analysis shows that  in spite of this ${\mathcal L}_{KG}$  should not be omitted. This term seems to {\em  play an important role for the coupling of matter and interaction fields  to gravity} (perhaps  only in cosmologically ``small'' regions\,?). 

In a first rough orientation, we may expect the derivations of $\phi$   at comparable orders of magnitude as $\phi$ itself. Because of the negligibility of ${\mathcal L}_{KG}$ for cosmology at large, we  conclude from comparison of $T^{(KG)}$ and $T^{(\Lambda )}$
\[  ||\gamma || \ll ||\rho _{crit} ||\; , \]
($|| \ldots ||$  the numerical value stripped from physical dimensions). Including dimensions $[\gamma ] = [E L^{-1}]$, the above estimation means
\beq  \gamma \ll  ||\rho _{crit} |\, [L^2] \; ,\eeq 
where $[\;\ldots\;]$ denotes physical dimensions and $E, L,T$ energy, length, time quantitities.

In order to find a bosonic mass for the vacuum field we 
 {\em  need not}  plug  an (external) mass term into the Klein-Gordon action of the vacuum field, as assumed by  Drechsler and Tann. They have argued that  physical gauge is due to an external mass term added to the Klein-Gordon action, and that all other masses are determined  by  the mass $m_0$ of the vacuum field, which serves as  a kind of ``measuring rod'' for energy and masses. If we take up their basic argument, but  apply it to the   spontaneously acquired mass $m_0$, we  find a {\em physical content for the observational gauge} and, sometimes together with it, the gauge defined above as Newton gauge.

In the Weyl geometric extension of general relativity mass $\tilde{m}$ and energy $\tilde{E}$ of particles or quanta have to be represented  by scalar quantities of weight $[[\tilde{m}]]=  [[\tilde{E}]]= -1$, i.e., as Weyl functions.   They stand in a fixed proportion to one another, in particular to the mass of the vacuum boson, 
\beq \frac{\tilde{m}}{m_0} = const = : m \; , \quad \quad \quad \tilde{m} = m \cdot  m_0 \; . \label{mass measurement}
\eeq 
Here $m$ is a scale invariant constant. 

If the spontaneous acquirement of mass according to (\ref{Klein-Gordon mass}) is of physical relevance,  material measurements will indicate  the scale invariant proportionality factor $m$ rather than $\tilde{m}$, which is its appropriate scale covariant  theoretical expression. The trajectory of particles is described by Dirac's scale covariant geodesics $\gamma _D(\tau )$  and the invariant mass factor $m$. The trace of scale invariant geodesics $\gamma _W(u) $ agrees with the trace of   $\gamma _D(\tau )$, and  mass scales along    $\gamma _W(u) $  like $\tilde{m}$. Therefore it is a matter of taste, or convenience, which description is chosen. In this context, we generally prefer the Dirac version $(\gamma _D(\tau), m)$.

Of course there is a preferred gauge, in which the mass of the vacuum boson $m_0$  is constant. By obvious reasons we call it the {\em constant vacuum boson energy} gauge, or simply {\em vacuum gauge}. 	It is compatible with Newton gauge in the absence of strongly varying matter fields, hence in the context of cosmic mean geometry.

 The spontaneously acquired mass of the vacuum field can also be used as a natural starting  point for Drechsler's  derivation of a {\em Higgs mechanism  coupled to gravity}  in the Weyl geometric frame \cite{Drechsler:Higgs}.  Drechsler proposed to characterize the vacuum structure by a 2-dimensional complex vector bundle $E \longrightarrow M$ with structure group $SU_2\times U(1)$. Then the vacuum field is extended from a complex scalar (purely gravitational) one to a  section of vanishing curvature  in $E$     of scale weight  $ -1$, i.e., a gravito-electroweak vacuum. Locally it appears as  a  ``scalar'' Weyl field $\Psi$ with values  in $\C^2$.    By a  local parallelization it   can be given  the form   
\[  \Psi ^{(0)} = \left(  \begin{array} { r }
 \Psi_1 \\ 
 \Psi_2 \\
 \end{array} \right)  \]
with $ \Psi_1 = const = 1$ and  $ \Psi_2 = \phi$.  In this way,  the electroweak  vacuum structure can be reduced to a complex scalar field in  Weyl geometry, which satisfies a Klein-Gordon equation as above. A ``reduction'', more precisely a 
 non-homomorphic mapping of the electroweak group action onto the stabilizer of $\Psi ^{(0)} $, leads to mass terms for other electroweak boson fields with 
non-vanishing curvature.  If this differential geometric version of the Higgs ``mechanism''  gives a physically correct picture of gravitational couplings of electroweak bosonic fields, we have $m_0 = m_H $, the Higgs mass. The experimental measurement of $m_H$ would allow to determine the value of $\gamma $. 

To complete the analysis of the Higgs mechanism,
Drechsler used his  additionally postulated (external)  mass term. We have seen that Weyl geometry offers   a  more natural self-gauging mechanism. This could not be seen by Drechsler, because  he passed over to the classical Riemann gauge shortly before the end of his analysis, like in (SCRG), and adopted the central hypothesis of traditional cosmology (\ref{central dogma}). In this way he needlessly gave away crucial advantages offered by  the Weyl geometric  extension of the theory. The two artificial restrictions (last minute Riemann gauge and external boson mass) can be avoided as indicated.

\section{Vacuum gauge and Weyl's gauge hypothesis \label{section vacuum gauge}}

Up to now we only dealt  with theoretically natural ways for gauge choices. This  opened up new thought possibilities beyond  the silent claim of semi-Riemannian  gravity (SCRG); but it remains still completely unclear whether it  leads to   new physical insight.  We now  look for physical specification of gauge. Here our attention is
mainly directed towards cosmology, but similar investigations could be brought to bear for other applications.

Drechsler's and Tann's proposal to consider the vacuum boson mass as the clue for observational gauge is very convincing. It deserves to be considered as a basic principle of scale covariant (Weyl geometric) gravity. \\[1mm]
{\bf Principle of vacuum gauge (VG).} {\em The vacuum boson mass serves as a kind of measuring rod for mass/energy in general and, with it (dualized), length, time etc. scales, i.e., observational gauge is given by vacuum gauge:
\beq M_0 = const \; \quad \mbox{in observational gauge}  \;\;  
\eeq 
}

 High precision measurements speak strongly against a varying gravitational constant \cite{Will:LivingReviews}. Similar to the standard approach in this respect,
we can therefore  explore how far we come with the following, simplest possible, assumption: \\[1mm]
{\bf Empirical postulate (EP).} {\em For the empirically accessible part of the universe the value of the gravitational constant  can be considered as observationally constant. We therefore assume that  in `empty' space (no strong matter fields above cosmic mean energy density) the observational gauge  of a Weyl geometric approach  coincides with Newton gauge, 
\beq G_N = const \iff \phi^{\ast} \phi = 1  \; \quad \mbox{in observational gauge. }  \;
 \label{empirical principle}
\eeq
 } 

Equation (\ref{Klein-Gordon mass}) shows that in the presence of strong matter fields (classical or quantum), which induce a varying scalar curvature $\overline{R}$, Newton gauge ($\phi^{\ast}\phi =1$) and vacuum gauge ($M_0 = const$) {\em cannot} coincide. That is different in the absence of strong fields. 

The principle (VG)  and  (EP) taken together presuppose the existence of a gauge (observational gauge)  in `empty space' for which the following holds:
\beq M_0 = const \iff \phi^{\ast}\phi =1 \iff G_N = const  \quad\;\; \label{hypothesis WGC} \eeq

Equ.  (\ref{Klein-Gordon mass}) shows that in such a  gauge also the scalar curvature $\overline{R}$ is scaled to a constant and coincides with Weyl gauge. Therefore in any Weyl geometric approach satisfying the VG hypothesis and (EP)
 {\em  Weyl gauge}   is {\em  of  particular importance}.
This gives  new  reasons to reconsider  an old hypothesis of H. Weyl, which looked rather ad hoc in its original version and has been  neglected for a long time by comprehensible reasons   \cite[298f.]{Weyl:RZM5}.\footnote{Much weaker, but already recognizable, in the 4th edition, translated into English \cite[308f.] {Weyl:RZMEnglish}}
 In order to make its methodological  status  as clear as possible, we state it in the following form:
{\lem In any scale invariant approach to cosmology  with Lagrangian density (\ref{total Lagrangian}), in which (VG)   and (EP) hold, 
 Newton gauge, vacuum gauge and Weyl gauge are identical,
\beq \phi^{\ast} \phi = 1  \iff M_0 = const \iff\overline{R} = const \; ,
\label{Weyl gauge principle} 
\eeq
and  coincide with observational gauge. \label{lemma Weyl gauge}}\\[1mm]

This is no ad hoc replacement  of the   the hidden hypothesis (SCRG) of the received approach. If  the comparison of observable masses at different places in the world is induced by their proportions to the vacuum boson and there is no external reason (varying mass field), which changes conditions, the Weyl gauge principle (\ref{Weyl gauge principle}) is due to a {\em self-gauging vacuum structure}. The latter serves as a kind of ``measuring rod'' in the sense of  Drechsler/Tann's analysis.

 If this assumption is physically realistic, it will be  consequential for  the coupling of interaction fields with gravity also inside massive systems, in particular for collapsed objects. In Riemann gauge and in Newton gauge scalar curvature  can acquire extremely high values near the critical boundary surface of a  black hole (the event horizon of a future trapped achronal set in Riemann gauge), if there is inflowing mass/energy. Length and time measuring units are  inversely  proportional to $M_0$ and indicate much larger volume  than in Riemann gauge. We  can thus expect much more volume in vacuum gauge close to the event horizon but still in the exterior. This volume expansion can be informally described as dark volume pockets  (``dark'', because the geodesic structure remains unchanged, only the Riemannian component of the metric  is modified). This may  lead to  different metrical and dynamical properties in  the environment of the classical event horizon than expected in the present theory. Under our assumptions, the singular behaviour of the
 semi-Riemannian structure may  be naturally dissolved by a more physical rescaling of the metric. In this sense,  a kind of physical ``resolution of singularities'' may take place for the singularity neighbourhoods which characterize galactic cores and quasars  in Riemannian geometry.\footnote{Seyfert galaxies seem to indicate a transition phase of galactic cores to active mass energy ejection, when the effectiv energy density has  sufficiently decreased by gravitational binding effects and  by  the conjectured metrical effects of vacuum gauge, to allow strong outflowing systems of trajectories (jets). }

\section{Cosmological mean geometry  \label{section cosmological mean geometry}}
For the description of cosmological mean geometry, i.e., under abstraction from local deviations by inhomogeneities of mass distribution, it makes sense to assume  the existence of a  (local or even global) fibration of 
space-time 
\[   \tau:  M  \longrightarrow T  \approx \R \; \]
with  spacelike fibres $S_{t} := \tau^{-1} (t), \; t \in T$, which are  maximally homogeneous and isotropic, and homothetic among each other. Then the Riemann gauged manifold $(M, (\tilde{g},0))$   can be isometrically characterized by a  warped product with warp function $a$
\[  M \cong  \R \otimes_a S_{\kappa }\; ,\]
where $S_{\kappa }$  is   a 3-dimensional space  of constant sectional curvature, 
$\kappa \in \R$. Topologically  we have $ M \approx \R \times S_{\kappa }  $, and $\tilde{g}$ can be given by a  
Robertson-Walker metric like in (\ref{Robertson Walker metric}). 

In this case, 
$ T ^{(em)} = T^{(KG)}=0 $, and  the r.h.s. of the Einstein equation reduces  to 
\beq   8 \pi \, G_N T :=     8 \pi \, G_N  T^{(m)} + T^{(\Lambda )} 
 \label{Robertson-Walker r.h.s.}   \eeq 
It is well known that here $T$ has the form  of  a fluid energy momentum tensor 
\[  T= (\mu + p) X^{\ast }\otimes X^{\ast} + p \, g  \; ,\]
with homogeneous and isotropic mass-energy density $\mu $ and pressure $p$, which can be expressed in terms of the warp function $a$ and its first two derivatives $a', a''$ \cite[346]{ONeill}.\footnote{$\mu = 3\, (8 \pi G_N)^{-1} (a'^2/a^2 + \kappa /a^2 )$, $p = - (8 \pi G_N)^{-1}(2 a''/a  + a'^2/a^2 + \kappa /a^2 )$ 
 \label{footnote Rob-Walker fluid}}
 It can be assimilated to equ. (\ref{Robertson-Walker r.h.s.}) in two ways,
\beqa
 T^{(m)} &=& T^{(fl)}, \hspace{17mm}  T^{(\Lambda )} = 0 \; ,\quad \mbox{fluid of pressure $p$, mass density $\mu$} ,\\
T^{(m)} &=& \rho _m X^{\ast} \otimes X^{\ast} \; , \quad T^{(\Lambda )} = \Lambda g  \; ,\quad \mbox{dust and $\Lambda $-term} \;\;  \\
 & & \hspace{20mm}
\quad \mbox{with $\rho _m = \mu + p$}
\quad \mbox{and} \; \Lambda = - 8 \pi G_N p \; ,
\eeqa
or a linear combination of both.

 Because of lemma \ref{lemma Weyl gauge}, we are particularly interested in Weyl gauged Robert\-son-Walker manifolds, respectively fluids. Under natural assumptions for the fluid, which will be analyzed in the next section (lemma \ref{lemma warp gauge case 2}), warp gauge will turn out as a physically interesting case. We thus have reasons to consider the class of Weyl geometric Robertson-Walker manifolds for which Weyl gauge and warp gauge coincide. This is a strong constraint. The following proposition shows that it reduces the infinite dimensional model class to a 
2-parameter subclass. The exemplars  of this subclass will be called {\em generalized Weyl universes}.

{ \prop  A Weyl geometric Robertson-Walker manifold with  warp function $a(\tau)$  (in Riemann gauge) characterizes a generalized Weyl universe $\iff$  the warp function  satisfies the differential equation
\beq a'' a + a'^2 = const  \label{WU DEQ} \; .
 \eeq } 
{\proof The scaling function  from Riemann gauge to warp gauge (\ref{warp gauge}) is $\Omega _W = \frac{1}{a}$.  Scalar curvature is of weight $-2$. Rescaling    from Riemann gauge to Weyl gauge (\ref{f gauge}) goes thus  by
$\Omega _{\overline{R}} = \overline{R}\, ^{\frac{1}{2}}$. 
With the expression for the scalar curvature of Robertson-Walker manifolds  \cite[345]{ONeill}
\beq  \frac{\overline{R}}{6}=  \left(\frac{a'}{a}  \right)^2 + \frac{\kappa }{a^2} +\frac{a''}{a} \label{scalar curvature Rob-Walker} \eeq
the claim of the lemma is a direct consequence of 
\[ \Omega _W \sim  \Omega _{\overline{R}} \quad \iff \quad \frac{1}{a^2} \sim (\frac{a'}{a})^2 + \frac{\kappa }{a^2} + \frac{a''}{a} \; .  \quad \quad  _{\boxempty }  \]
 }

For the Weyl geometric approach, the role of this differential equation is comparable to  the Friedmann-Lemaitre equation for  classical (semi-Rie\-mann\-ian) cosmology. 
It  specifies  a physically reasonable 2-parameter subclass of all Weylian Robertson-Walker solutions (for a given r.h.s $const$). 

The two-parameter set of solutions   of the non-linear differential equation (\ref{WU DEQ}) is surprisingly simple.   For $H^2 := const$  it is 
\beq  a(\tau) = \pm  \sqrt{  (H\, \tau ) ^2  + C_1 \tau  + C_2  }  \; ,  \label{solution DEQ}
\eeq
with constants $C_1, C_2 \in \R$. 

Clearly the solution class is much simpler than the Friedmann-Lemaitre class. It even contains a distinguished special case. 
For large values of $\tau$, the general solution rapidly approximates  the special solution  $a_0$ with $C_1 =C_2 = 0$  (uniformly on the remaining interval $(\tau , \infty )$ and with rapid convergence $|(a - a_0)_{| (\tau,\infty )}|
 \longrightarrow 0$ for ${x\to\infty}$) 
\beq a(\tau) \quad  \longrightarrow  \quad a_0 (\tau) := H \, \tau  \; . 
 \eeq
In this sense, the linear warp function $a_0(\tau) = H \, \tau$ characterizes an attractor solution of the warp function for the whole class of generalized Weyl universes. 

In order to pass to the warp gauge, we   have to reparametrize the time cordinate $t := H^{-1} \log H \tau \iff \tau = H^{-1} e^{H t}$. Then the Robertson-Walker metric with linear warp function $ds^2 = - d\tau^2 + (H\tau)^2 d\sigma_{\kappa } ^2$ acquires  a ``scale expanding'' form (Masreliez)
\[ \tilde{g}: \quad ds^2 =  e^{2H   t } (- d t+ d\sigma_{\kappa }^2  ) \]
and can be rescaled by $\Omega = e^{-Ht}$ to 
\beq   g: \quad ds^2 = - d t ^2 + d \sigma_{\kappa } ^2 \, ,\quad \quad  \varphi = H d t    \label{Weyl universe}\eeq 

These models have been called  (special) {\em Weyl universes}  elsewhere \cite{Scholz:ModelBuilding}, and so they will be  here. Basically  they can be  derived from the old  static models of cosmology by superimposing  a  time-homogenous scale connection. That results in a small deformation of the metrical structure.

If the general Weyl universe with warp function (\ref{solution DEQ}) is treated similarly, the Hubble gauge becomes
\[  \varphi =   \frac{H^2 + 2C_1 H e^{-Ht} + (2C_1^2 - 4 C_2 )e^{-2Ht}  }{H + 2 C_1 e^{-Ht} + 4 C_2 e^{-2Ht} }dt \approx H \, dt \quad \mbox{for ``large'' }  \; t \, .  \]
The rapid approximation behaviour stated above can easily be read off from this explicit expression.

We resume our result as: 
{\thm The (special)  Weyl universes (\ref{Weyl universe})  with curvature parameter $\kappa $  are stable solutions of (\ref{WU DEQ}) in the class of generalized Weyl universes. Any other member of the class is a  generalized Weyl universe  of the same $\kappa $ and  approximates, for increasing $t$, the special Weyl universe  in fast approximation (in the $sup$ norm).  }\\[0.5mm]

 Another result is not difficult to prove.  Here it will be stated without proof, because its   more mathematical in nature than the rest of this contribution.

{\thm  Up to isomorphism, (special) Weyl universes form a 
1-parameter family of models  with the only essential  metrical parameter (module)
\beq  \zeta := \frac{\kappa }{H^2} \; .
\eeq }

The Hubble constant (here $H = H_0 c^{-1}$)  itself is no structural parameter of the model class. 
It   remains, of course, a clue  for the adaptation of the model to observational data.

We set 
\beq \Omega := \frac{\mu }{\rho _{crit}} \; ,
\eeq
with 
\[   \rho _{crit} = \frac{3 H^2 \, (\phi^{\ast} \phi)}{8 \pi g_N} [c^4] =   \frac{3 H^2 }{8 \pi G_N} [c^4]  \]  
 the   
 critical density, as usual. Then 
the total net energy density,  $\Omega = \Omega _m + \Omega _{\Lambda } $, determines the geometrical module $\zeta $ by the condition \cite{Scholz:ModelBuilding}
\beq  \Omega = \zeta +1 \; . \label{Omega zeta}
\eeq
This relation is the Weyl geometric analogue of the  well known  balance  for the Friedmann-Lemaitre class,  $\Omega _m + \Omega _{\Lambda } + \Omega _{\kappa } = 1$. Here the other relative densities are 
\[  \Omega _m = \frac{2}{3} \Omega ; , \quad \Omega _{\Lambda } = \frac{\Omega }{3}.  \]

Equ. (\ref{scalar curvature Rob-Walker}) shows
$  \overline{R} = 6 (\zeta +1)H^2$. 
Furthermore 
\[   \Lambda = \beta (\phi^{\ast} \phi ) = 8 \pi g_N \,  (\phi^{\ast} \phi )^{-1} \, \rho _{\Lambda } = 3 H^2\, \Omega _{\Lambda } = H^2 (\zeta +1) \; .
\]
 For Weyl universes the crucial r.h.s. factor in (\ref{Klein-Gordon mass}) in observational gauge ($=$ Weyl gauge) becomes
\beq \frac{\overline{R}}{2} - 2 \beta (\phi^{\ast}\phi )  = H^2 (\zeta +1) = \kappa + H^2 \; .
 \eeq 
That is exactly 
the principal sectional curvature in the spacelike fibres of the Riemann gauge  (in the warped product!) \cite[345]{ONeill}. The mass of the vacuum boson in a Weyl universe with module $\zeta $ is   now given (in Weyl gauge) by
\beq  
m_0^2 c^4 = \left(  \frac{\hbar c}{g_N} \right)  \frac{c\, \hbar}{8 \pi \gamma } H^2 (\zeta +1) \; ,
\label{vacuum boson mass Weyl universe} \eeq 
where $  \frac{\hbar c}{g_N}    = E_{Pl}^2 $ is the squared Planck energy.

\section{Fluid mass term \label{section mass term}}
Usually  a perfect fluid {\em Ansatz} is chosen to describe  the mass term in cosmological solutions of the Einstein equation, e.g.,   \cite[69f.]{Hawking/Ellis}. It can easily be adapted to the Weyl geometric context. Consider a matter Lagrangian density 
\beq {\mathcal L}_m = - 2 \rho (1+\epsilon ) \sqrt{|g|} \; , \label{fluid Ansatz}
\eeq
where $\rho $ is a (real) scalar field which represents the  energy density. At the moment we do not analyze in which gauge it becomes constant; for special cases  that  will become clear later (case 2). Its scale weight is  $[[\rho ]] = -4$, in agreement  with dimensional conventions for mass/energy density.  $\epsilon $ is a real scalar function  of $\rho $, characterizing the proportion of {\em internal energy}  to $ \rho$. Obviously its gauge weight has to be $[[\epsilon ]]=0$.  $ {\mathcal L}_m$  is a scale invariant density.

In normal fluids $\epsilon $ represents  the elastic potential and increases monotonically with $\rho $. Following \cite{Fahr:BindingEnergy} in the search for a mathematical characterization of  gravitational self binding energy, we investigate whether   $\epsilon $ may be employed as an expression for the latter (at first only formally). Increasing mass energy density $\rho $ ought to lead to higher gravitational self binding energy which  has to be subtracted from the original mass energy. That reduces the internal energy of the fluid. Different to ordinary fluids, we therefore expect here 
\beq  \frac{d\epsilon }{d \rho } \leq 0  \quad \quad \quad \mbox{with equality   only for $\epsilon  = 0$}\; . \label{negative self-binding energy} \eeq
The  effective (``net'') {\em total energy density} of the fluid  is
\beq \mu = \rho (1+\epsilon ) \; . \label{effective energy density}
\eeq

Let $X = (X^{\mu })$ be a timelike unit Weyl vector field of  weight $[[X]]=-1$ and 
\[  j := \rho X \]
the respective mass energy current. If the condition of   conserved  current
\beq div\;  j = D_{\mu }  j^{\mu } = 0\label{conserved current}
\eeq 
is satisfied, the {\em  pressure} of the fluid is 
\beq p = \rho ^2 \frac{d \epsilon }{d \rho } \; .\label{pressure}
\eeq

Variation of the metric in Riemann gauge under the restriction (\ref{conserved current}) leads to the energy stress tensor of the fluid \cite[70]{Hawking/Ellis}
\beq T^{(fl)}_{\mu \nu } =  - \frac{1}{\sqrt{|g|} } \frac{\delta {\mathcal L}_{m}}{\delta g^{\mu \nu }}  = \rho _{m} X_{\mu} X_{\nu } + p \,g_{\mu \nu } \; , \label{fluid energy tensor}
\eeq
with 
\beq  \rho _m := \mu +p 
\eeq
its net mass energy density. The fluid is  {\em isentropic}, if  $p$ is function of $\mu $ only. According to \cite{Hawking/Ellis} this is a sufficient condition for $T^{(fl)}$ being derivable from a Lagrangian.

All  expressions in the above derivation  are scale covariant. $T^{(fl)}$ is a Weyl tensor field of weight $-2$ and therefore compatible with (\ref{general matter tensor}), although it has been derived from a slightly more constrained variational principle. Therefore it seems  justified to transfer the fluid Ansatz  to scale covariant gravity.

Two cases are of particular interest for us.\\[0.5mm]
\noindent
{\bf  Case 1}:  {\em Ordinary dust.}\\
 No pressure, no internal energy
$\epsilon =0, \; p = 0, \; \mu =\rho = \rho _m$. \\
Lagrangian density
\[  {\mathcal L}_m = - 2 \rho _m \sqrt{|g|} \; ,\]
mass energy  tensor 
\[  T^{(m)} = T^{(fl)}=\rho_m X^{\ast} \otimes X^{\ast} \;  \]
($X^{\ast}$ the $g$ dual of $X$, cf. below).\\[0.5mm]

\noindent
{\bf  Case 2}:  {\em Dust with  self binding energy given by  $\epsilon $, coupled to the $\Lambda $-term.}\\
In the Lagrange density (\ref{fluid Ansatz})
we assume a negative self binding energy given by $\epsilon $, such that the pressure (\ref{pressure}) is gauge covariant with weight $[[p]]=-4$, like $\rho $. That implies $\frac{d \epsilon }{d\rho } \sim \rho ^{-1}$. We therefore  assume a logarithmic self binding coefficient of the form 

\beq  \epsilon = -\log \left(\frac{\rho }{\rho _{crit}} \right)^{ \alpha  } = - \alpha  \log \tilde{\Omega } \label{self binding Ansatz}
\eeq
with  $\rho >0$, and $ \alpha  > 0$ a logarithmic power characteristic for  the self binding energy. (Confusion of this coefficient $\alpha $ with the one from the Lagrangian is excluded because of the  substitution (\ref{alpha substitution}).) 
Here we use the abbreviation
\beq  \tilde{\Omega } := \frac{\rho }{\rho _{crit}} \; .
\eeq

  $\frac{d\epsilon }{d\rho }= - \frac{\alpha }{\rho }$ is negative, as we expect for gravitational self binding energy.  Under  condition  (\ref{self binding Ansatz}), the pressure term becomes proportional to the original energy density:
\beq p = \rho ^2 \frac{d\epsilon }{d\rho } = - \alpha \rho  \;  \label{pressure case 2}
\eeq
The {\em net total energy} density is
\beq  \mu = \rho (1+\epsilon ) = \rho (1- \alpha \log \tilde{\Omega }) \; .\label{total energy density case 2}
\eeq
 $\rho $ will be called {\em ``gross'' energy density} in the sequel.

The energy stress tensor of the fluid decomposes into two terms
\[  T^{(fl)} = (\mu + p) X^{\ast }\otimes X^{\ast} + p \, g  \; ,\]
with $ X^{\ast} = (X_{\mu })$ the dual of $X=(X^{\mu })$ with respect to the Riemannian component of any gauge and  $[[X^{\ast}]]=1$. The  contraction is $\complement (X \otimes X^{\ast}) = X^{\mu }X_{\mu }  = g(X,X)=- 1$.
$T^{(fl)}$ has  the form of an ordinary dust energy tensor and a cosmological term: 
\beqa   8 \pi G_N \,  T^{(fl)} &=&  8 \pi G_N \, 
  T^{(m)}  + \; T^{(\Lambda )} \; , \\
  T^{(m)} &=& \rho _m   X^{\ast}\otimes  X^{\ast} \, , \quad   T^{(\Lambda )} =  8 \pi G_N \,  p\, g = -  8 \pi G_N \,  \rho_{\Lambda } g \; ,
 \eeqa
with $p= - \alpha \rho =-\rho _{\Lambda } $ and 
\beqarr  \rho _m  &=& \mu  - \alpha  \rho  = \rho (1- \alpha  (1+ \log \tilde{\Omega })) \; \label{fluid Ansatz matter tensor}\\
\rho_{\Lambda }  &=&  \frac{\Lambda }{8 \pi \, G_N }   \;, \quad \Lambda = 8 \pi g_N (\phi^{\ast} \phi)^{-1} \alpha \rho   .  \label{coupled Lambda}
\eeqarr
 At first sight it may look like a  formal trick that a ``cosmological'' term drops out of the isentropic fluid Ansatz. But  {\em any} mathematical descriptor for gravitational self energy ought to be a kind of cosmological correction to the ordinary relativistic r.h.s. of the Einstein equation. In  this sense the  metaphor of  cosmic matter and 
``ether'' (in the language of Einstein, Weyl e.a. in the 1920s) as a kind of fluid with negative pressure \cite[359, 429]{Levi-Civita:1926} may  perhaps  serve as a  fruitful heuristics for  understanding  the $\Lambda $-term  still today.
We have to keep in mind, however, that  the $\Lambda $-term plays a distinctive role in the variation. 
 $ {\mathcal L}_{\Lambda } $ is varied with respect to general $g$ and  ${\mathcal L}_m $ under the constraint (\ref{conserved current}) only.

Scale gauge weights  of all terms are consistent with what we demand for the r.h.s. tensors of equation (\ref{Einstein equation}). 

Contraction of the l.h.s of the $(1,1)$ Einstein equation leads to the well known term
\[ \overline{R} - 4 \frac{\overline{R}}{2} = - \overline{R}  \; .\]
Contraction of the $(1,1)$ raised energy stress tensor at the r.h.s. gives, up to the constant $ 8  \pi  G_N  $,
\beqa  \left(  \complement (\rho _m X \otimes X^{\ast}) +4 p\right)  = - \rho _m + 4p    
& = & -   \mu + 3p =  \rho \, (3\alpha -1-\epsilon )
\eeqa
The l.h.s. is constant in Weyl gauge. Differentiation of the r.h.s. shows that the latter is constant, if $\rho '=0$ or $ \epsilon = 1-2\alpha $, which leads back to $\rho =const$. By (\ref{pressure case 2}), (\ref{total energy density case 2})
this implies  constant effective dust energy density $\rho _m$ and  constant negative pressure $p$ and $\Lambda $ in Weyl gauge. According to  (\ref{hypothesis WGC}) this means that {\em   in observational gauge} (which is equal to Weyl gauge by lemma 1) 
\[ \rho _m = const \; , \quad \quad p = const \; ,
\]
 independently of the choice of $\alpha $. 
For Robertson-Walker fluids, i.e., fluids with the same type of symmetry as Robertson-Walker manifolds, we draw the consequence:

{\lem   Weyl gauge and warp gauge coincide for case 2 fluids. \label{lemma warp gauge case 2}}

{\proof If Weyl gauge $(g_W, \varphi_W )$ has a Riemannnian component
\[ g_W: \quad ds^2 = - dt^2 + a_W(t)^2 d\sigma _W^2 \; ,  \]
the   (rest) warp function $a_W $ must be constant. Otherwise the conserved current condition (\ref{conserved current}) would imply  a non-constant $\rho _m$, in contradiction to what we  have just seen. Thus Weyl gauge is the same as warp gauge. $_{\openbox}$}\\[0.2mm]

This property of case 2 fluids has striking consequences for the physical geometry generated by them. In the last section it has been shown that it restricts the geometry to generalized Weyl universes and then, by theorem 1,  to a (special) Weyl universe geometry as a stable  attractor which is approximated in fast convergence.

\section{Coupling of the vacuum field to  mass energy \label{section mass term}}
A most  interesting property of case 2  is its inherent coupling of  the vacuum tensor to mass energy. H.-J. Fahr and M. Heyl  have remarked  that present cosmology  presupposes a  cosmic  vacuum which acts on matter and geometry, without itself  being  acted upon by the latter \cite{Fahr:BindingEnergy}. This  unsatisfactory state of affairs  may be improved by investigating  possible  coupling conditions of  $T^{(\Lambda )}$ to  matter  $T^{(m)}$. In such a case  $\Lambda$ can no longer be a true constant. While this may seem irritating, at first glance,  for the received perspective,   it is  no problem  in  the Weyl geometric program. Here  we have
$ \Lambda = \beta (\phi^{\ast} \phi) $ 
by equation  (\ref{Lambda and beta})   anyhow,  and the question turns into the one, whether the vacuum field $\phi$  may  couple to   matter density averaged over sufficiently large cosmological regions in a physically reasonable way.

Fahr and Heyl argue  that an increasing vacuum energy density should lead to an increasing flow of particles from the sea of  vacuum fluctuations  to the matter component of $T$, i.e., matter creation. The other way round,  an  increase of matter density should lead to a rising gravitational self binding energy. This results in a transfer of energy from (net) gravitating matter to the gravitational field, and with it to the vacuum component of $T$.

These qualitatively  convincing arguments are difficult to quantify because of the problem of an undefined energy of the gravitational field in general relativity and the lack of tested knowledge of matter creation from the vacuum. Fahr/Heyl propose an interesting method which explores this field; but  we cannot be sure that their route will  lead to a valid  answer. We take up   the question and their general line of approach, but explore the latter in  the perspective of scale covariant gravity.

In case 2  above the total (net) energy density $\mu = \rho +\epsilon $ has been  reduced in comparison with the original  (gross) mass energy density $\rho $ by $\epsilon = - \log \tilde{\Omega }^{\alpha }$ because of overall gravitational binding effects. Moreover, a part of the total energy density, $\alpha \rho $, is transferred from the dust (mass) term $T^{(m)}$  to $T^{(\Lambda )}$. Under the assumption of  (\ref{self binding Ansatz})  the term $T^{(\Lambda )}$ contains the observable energy momentum of the gravitational self binding effects. Similarly  {\em any negative energy fluid Ansatz (\ref{fluid Ansatz}), (\ref{negative self-binding energy})  should give a natural coupling of the vacuum tensor to  mass energy} comparable  to 
(\ref{coupled Lambda}).

This shows that it is not at all unreasonable to expect links between partial Lagrangians. The link in our case 2 example can be described by comparing it with 
 a pure  dust matter Lagrangian (case 1)
\[  {\mathcal L}_m = - 2 \rho _m \sqrt{|g|} \; \quad \rho _m =  \rho (1-\alpha (1+\log \tilde{\Omega })) \]
and a cosmological term
related  to it  (cf. (\ref{coupled Lambda}))  by the condition
\beq  \beta = \frac{8 \pi g_N }{(\phi^{\ast} \phi )^2} \alpha \rho  \; , \quad \i.e. \quad  {\mathcal L}_{\Lambda } = -  16 \pi g_N \, \alpha  \rho  \sqrt{|g|} \; .  \label{central coupling condition} \eeq
Clearly  the {\em self binding exponent  $\alpha $ is here the crucial parameter} for the  balance between  mass energy density and the vacuum tensor.

 In order to acquire a better understanding of structural possibilities for cosmological models,  it may be useful  to investigate such  coupling conditions more generally,  even  independent of  the special background theory  of isentropic fluids, case 2.  The essential point is a coupling of type (\ref{central coupling condition}).
More generally we know from equ. (\ref{Klein-Gordon mass}) that
\[ \Lambda = \beta (\phi^{\ast} \phi ) = \frac{\overline{R}}{4} - \frac{\gamma }{2 \alpha } M_0^2 \; .\]
 In vacuum gauge it rises with $\overline{R}$, and thus with mass energy density. 
 To concentrate ideas we continue here, however, by referring to the case 2 example.

\section{Equilibrium condition \label{section equlibrium}}
For a classical Robertson-Walker perfect fluid, i.e., a solution of the 
semi-Rie\-mann\-ian  Einstein equation with r.h.s  tensor (\ref{fluid energy tensor}), the expansion dynamics is regulated by Raychaudhuri's  differential equation for the warp (expansion) function $a(t)$ \cite[346]{ONeill}
\beq \frac{a''}{a} = - \frac{4\pi g_N}{3} (\mu  + 3p ) \; . \label{Friedmann equation}
 \eeq 
This leads to the known equilibrium condition for the Einstein universe and other non-evolving (``static'') Robertson-Walker solutions  with constant sectional curvature of the space sections:
\beq  3 p_0 = -  \mu \label{hyle condition}
\eeq 
For $p<p_0$, the gravitational forces of mass energy win over the negative pressure,  and the corresponding Robertson-Walker cosmology will collaps  in the long run. For $p>p_0$ it will expand \cite[A44]{Ellis:83years}. The balance equ. (\ref{hyle condition})  shall be called the {\em hyle condition} of a cosmic fluid. It is satisfied for
 Robertson-Walker fluids  $\iff$ $a'' = 0\,$; i.e., not  only in the static case (cf. footnote (\ref{footnote Rob-Walker fluid})).

Clearly this equilibrium is attained for case 2 (cf. (\ref{pressure case 2}), (\ref{total energy density case 2})), iff
\[ 3 \alpha  \rho   = \mu =\rho (1+\epsilon ) = \rho (1 - \alpha  \log \tilde{\Omega } )\; . \]
This condition establishes  a  peculiar relationship between the (logarithmic)  binding energy exponent $\alpha $ and the equilibrium value $\tilde{\Omega }_0$ of (relative) gross energy density:
 \beq \alpha  = (3 + \log \tilde{\Omega}_0)^{-1}  \quad \quad \Longleftrightarrow \quad \quad   \tilde{\Omega }_0 = e^{\alpha ^{-1} -3}\; \label{case 2 equilibrium}
\eeq

In order to understand better what its ``peculiarity'' consists of, 
we consider  small deviations $h$ of gross energy density from the hyle condition $\rho _0= \tilde{\Omega }_0 \rho _{crit}$,
\[  \rho = \rho _0 + h \;  .\]
The self binding energy will then be shifted  from the equilibrium value $ \epsilon (\rho _0)$ to 
$\epsilon (\rho ) = - \alpha  \log \tilde{\Omega } $ 
and the pressure from $p_0 = -   \alpha  \rho_0$ to 
\[ p(\rho ) =   -  \alpha  (\rho _0 +h) \; . \]
The difference of the l.h.s. of the hyle condition is  
\[  \Delta (3p) =  -  3\alpha  h \]
The   change of total (net) energy density close to the hyle (equilibrium) value $\mu _0 = \rho _0 (1- \alpha  \tilde{\Omega }_0)$ 
is
\[  \Delta \mu = h \cdot \frac{d \mu }{d \rho }_{|\rho _o } + o(h^2) \; .\]
Because of
\[ \frac{d \mu }{d \rho } = 1-  \alpha  (1+ \log \tilde{\Omega })  \]
we find
\[ \frac{d \mu }{d \rho }_{|\rho _o }  = 1-  \alpha  (1+ \alpha ^{-1}
-3) = 2 \alpha   \; \]
and thus
\beq   |  \Delta (3p) | > \Delta \mu \quad \mbox{for small } \;\; h\neq 0 \; . \label{attractor condition}  \eeq
For  $h>0$  the increase of negative pressure of  $T^{(\Lambda )}$ dominates over the contracting effect of  increasing  gross energy density   (at least in the regime where the  linearized estimation is applicable). The originally increased mean density $\rho $ is forced back towards the hyle condition $\rho _0$. If it passes to the other side,  $h<0$, the effect of (\ref{attractor condition}) works the other way round; $\rho _0$ turns out to be a stable equilibrium point. 

We  formulate our conclusion as
{\prop   Under the assumptions of case 2,  the hyle cosmology (\ref{hyle condition}) characterizes  a  stable equilibrium for  a   Robertson-Walker fluid with given self binding energy exponent $\alpha $, or the neutral state of an  oscillatory mode.}\\[1mm]
 If oscillations are damped by intrinsic reasons the hyle point is  a 
(local) attractor. 
Notice that in the classical picture it need not be static. Its generic state is a uniform ``expansion'' with $a''=0$, i.e., a linear warp function
\[  a (\tau )  = H \, \tau \; , \quad H = const \;  \]
and characterizes a Weyl universe (section \ref{section cosmological mean geometry}).

Even if we  consider  case 2 as a predominantly methodological tool   for investigating possible effects of matter  vacuum coupling,  we should not consider it as  a mere ``toy'' example. The existence of a stable attractor may   be  characteristic for the model space of other, more general, gravitational binding Ansaetze for ${\mathcal L}_{\Lambda }$.
It seems worthwhile to investigate which other self binding assumptions lead to comparable stability effects. From the mathematical point of view, it would be illuminating to know the maximal subspace of all Robertson-Walker fluids, in which the hyle cases are stable. Theorem 1 teaches us that  generalized Weyl universes belong to it.

Stable solutions  have become doubtful since Eddington's counter argument against Einstein's first cosmological model \cite{Eddington:Instability}. The rejection has become generally accepted   (Einstein included), after the interpretation of cosmological redshift as an effect of space expansion became dominant. It  was even further supported by the proofs   of the general singularity theorems by R. Pensore and S. Hawking in the 1960s.\footnote{For a crystal clear overview of the state of the arts, see \cite{Ellis:83years}.} {\em They do not  apply, however, in situations where  a coupling of ${\mathcal L}_m$ and ${\mathcal L}_{\Lambda }$ stabilizes the structure.} Our case 2 example shows that   the convergence of past directed timelike unit vector fields, which is one of the necessary premisses in the singularity theorems 2 to 4 of \cite{Hawking/Ellis}, can be excluded for specific, not even particularly strange coupling conditions.

\section{Cause of cosmological redshift and the Mach principle  \label{section Weyl universes}}
The  expanding space paradigm for the explanation of cosmological redshift does not admit  equilibrium  cosmologies. The last passage  makes sense  in a Weyl geometric context only. At other occasions it has  been  discussed, how well integrable Weyl geometry (IWG) suits the task  of analyzing different physical assumptions for the causes of cosmological redshift from a common mathematical vantage point \cite{Scholz:Extended_Frame}. Different scale gauges allow to transform one and the same cosmological redshift function $z(\tau )$ ($\tau $  the cosmological time parameter) into different mathematically equivalent expressions. IWG is  the appropriate framework for comparing  the expanding space interpretation of $z$ by  the warp function of Robertson-Walker models (\ref{Robertson Walker metric}), 
\[ z(\tau ) = \frac{a(\tau )}{a(\tau _0)}  - 1\; ,\]
  with an energy scaling effect for photons of the Weylian length (scale) connection
\[ z(\tau ) = e^{\int_{\tau _0}^\tau  \varphi (c ' (u)) d u  } \; ,
\]
or with  a combination of both ($c(u)$ differentiable path). 

If in observational gauge, the physically most relevant  one,   the warp function is completely gauged away, the cause of cosmological redshift can no longer be seeen in a real ``expansion'' of space sections. It rather appears to be a purely field theoretic effect of the interaction of photons (the Maxwell field) and the gravitational vacuum field $\phi$. Cosmological redshift is then a ``higher order'' effect of gravity, in the sense of remaining unexpressed in the semi-Riemannian approximation of the theory. Its simplest mathematical expression is given by the Weylian scale connection, which appears here as the  {\em Hubble connection} $\varphi$.

Lemma \ref{lemma warp gauge case 2} and theorem 1 imply that the physical geometry established by case 2 Robertson-Walker fluids is very well approximated by a Weyl universe. Its crucial parameter (mathematically $\zeta $, physically $\Omega = \zeta +1$)
 is determined by the logarithmic exponential $\alpha $ of the gravitational self binding coefficient $\epsilon $. We know from equs.
 (\ref{total energy density case 2}) and  (\ref{case 2 equilibrium}) that 
\beq \Omega = \Omega _0 = \frac{\mu _0}{\rho _{crit}} = \tilde{\Omega }_0 (1-\alpha  \log \tilde{\Omega }_0) = 3 \alpha  e^{\alpha ^{-1}-3} \; .\eeq
Therefore the  equilibrium value of the geometric module $\zeta $ for case 2 fluids is 
\beq \zeta = 3 \alpha  e^{\alpha ^{-1}-3} -1 \; . \label{zeta alpha}
\eeq 

$\alpha  = \frac{1}{3}$ corresponds to a spatially flat Weyl universe ($\zeta = \kappa =0$,  Minkowski-Weyl universe), $   \frac{1}{3} < \alpha < 1 $ characterizes  a subset of  the Lobachevsky-Weyl case ($ 0 > \zeta \gtrapprox  - 0.6$, for which $\kappa <0$), and  $0< \alpha  < \frac{1}{3}$  Einstein-Weyl universes with positive spatial curvature, $\zeta > 0 \Leftrightarrow \kappa >0$.

 Equ. (\ref{zeta alpha}) constrains total  energy density to  $ \Omega  \gtrapprox 0.41$. i.e., $\Omega _m \gtrapprox 0.27$. Therefore a massless case 2 fluid Weyl universe,$\Omega =0$, does not exist, while the space-time of special relativity appears  as a Minkowski-Weyl universe with  neglected Hubble connection. It corresponds to $\Omega =1$, i.e., critical total energy density. In this sense, our model class of Weyl universes  has an inbuilt implementation of Mach's principle. {\em  Even a  Machian ``cause'' for the inertial structure of Minkowski space becomes apparent}: A  total net energy density $\mu = \rho _{crit}$, corresponding to a logarithmic self energy exponent $\alpha  = \frac{1}{3}$, while the Hubble connection has been ``forgotten'', i.e., has been abstracted from.  

Although our case 2 assumptions were derived from formal considerations (scale covariance conditions for $\rho $ and $p$ of isentropic fluids), our analysis shows a striking (theoretical) convergence of stepwise refined conditions toward the class of Weyl universes.  They have  apparently  such  extraordinary mathematical properties  that structurally similar, but more general assumptions for the  coupling of the vacuum term to mass  may be expected to  lead to the same model class. It is tempting to  conjecture that a purely gravitational self binding energy of electrically neutral matter belongs to the logarithmic exponential $\alpha = \frac{1}{3}$, corresponding to the Minkowski-Weyl solution. Partial ionization of the diluted matter content (mainly hydrogen) may be responsible for  reducing  the value. Observational values of SNIa etc. indicate a good empirical behaviour for $\alpha \approx 0.21$.

\section{A short look at observational data and conclusion \label{section observational data}}

 {\em Einstein Weyl universes} with positive spacelike curvature $\kappa > 0$ are observationally clearly  distinguished.
The actual data of  supernovae luminosities show a perfect fit  of the Hubble diagram of Weyl universes with positive sectional curvature to the supernovae observations. Weyl universes of 
$ \zeta \approx 2.6 $   give the best fit with the most recent SNIa data \cite{Riess_ea:2007}. The estimated confidence interval is
\beq  \zeta  \in [2.19, \; 3.0]  \label{zeta confidence interval} \; . \eeq

Supernovae data are fitted with  mean square error $\sigma_{W} \approx 0.21$ (at $\zeta = 2.6$), which lies {\em below} the mean square error of the data set $\sigma _{dat} \approx 0.24$ .
The fit quality is better than for  the Friedmann-Lemaitre model class which leads to a mean square model error $\sigma _{FL} \approx 0.27$, with $ \Delta \sigma \approx  0.03$ {\em above} the data error and twice as much above  $\sigma_{W} $. This difference is, however, still far from significant ($\sigma _{FL}-\sigma _W  \approx 0.2\, \sigma _{dat} $). We shall have to see what happens when  measurements to higher redshift values or of higher measurement precision become available \cite{Kowalski_ea}.   For $0 \leq z \leq 1.2$ the Hubble diagrams in the two models differ only very little. After an intersection of the two curves close to  $z  \approx  1.2$, they  have a noticable and increasing difference. (The standard model predicts a more rapid decrease of luminosities.) With sufficiently far and/or precise measurements of SNI beyond $z\approx 1.2$, we  expect a   statistically significant discrimination criterion between the two model classes.

The Einstein-Weyl geometry with  $\zeta \approx 2.6$ indicates relative energy densities $\Omega _m \approx 2.4$ and $\Omega _{\Lambda } \approx 1.2$. Before we draw  rash conclusions from the  unexpected high mass density values, we have to realize that the upper limit of baryonic mass density of the standard approach is no longer valid in  equilibrium cosmology  (from this vantage point, primordial nucleosynthesis appears as a theoretical artefact of the standard approach).  A much higher amount of {\em classical} matter,  molecular hydrogen and low energy hydrogen plasma, both distributed over large  inter cluster regions, astronomically extremely difficult to trace and unobservable by the dynamical method of cluster dynamics, may very well explain such a high value. The homogeneity of the mass distribution could even be much higher than estimated at present, because   clusters and super clusters  would lose their predominance as tracers for dark matter \cite{Peebles:Probing}.

With  $ \zeta \approx 2.6 $, the energy of the vacuum boson  can  be determined up to the unknown coupling constant $\gamma $ or up to 
\[ \tilde{\gamma }:=    \frac{8 \pi}{c \hbar} \gamma \;   \]
by equ. (\ref{vacuum boson mass Weyl universe}),
\beq  (m_0\, c^2)^2 \,  \tilde{\gamma } = E_{Pl}^2 H_1^2 (\zeta +1) \approx 3 \; eV^2 \, cm^{-2} \, . \eeq
If vacuum gauge still holds for  high energy experiments,  the presently expected order of magnitude for the Higgs boson, $m_H \sim 10^{11}\, eV$, would imply
\[  \tilde{\gamma }\sim 10^{-22} \, cm^{-2} \; , \quad \quad    \gamma \sim 10^{-27} eV cm^{-1} \; .\]
But it is unclear, whether  vacuum gauge remains applicable  under these 
 conditions.\footnote{If so, it could indicate an additional coupling of $\phi$ to quantum matter fields or of a non-constant $G_N$. 
}

Some of the other empirical data prefer Weyl  universes more clearly (quasar magnitudes, quasar frequencies,  Pioneer anomaly),\footnote{Cf.  \cite{Scholz:Pioneer,Scholz:ModelBuilding}; an updated discussion of present data is in preparation.} while still other observational evidence speaks  in favour of the expanding space cosmology, or may speak in favour of it in the near future (most importantly, but still undecisively star formation rate). Contrary to a general creed, neither the exact Planck characteristic of the  microwave background nor its anisotropy structure can serve as a differentiating criterion between the expanding and the equilibrium paradigms. They go well in hand with both. A  clear discrimination between the different approaches may rather be expected from the antenna systems,   designed for tracing the conjectured $21\;cm$ line of non-ionized monatomic hydrogen at redshifts above $z \approx 7$. If the latter  will be observed  within the next 5 to 10 years  with intensity significantly above average, the standard conjecture of a  reionization period for cosmic hydrogen, somewhere in the interval  $7< z < 30$,  will be empirically confirmed.  That would be a decisive empirical triumph over most (all?) competing approaches.  {\em If the experiment turns out null,  the outcome will be   converse. }

In our context of equilibrium cosmologies of case 2 fluids, 
 the present  curvature values (\ref{zeta confidence interval}) determined by supernovae data indicate the following value for the  hypothetical binding energy exponential (\ref{zeta alpha}):
\beq  \alpha  \in [0.205, \; 0.218] \quad \mbox{or } \;\;   \alpha   \approx 0.211 \pm 0.07
\eeq
We have to leave it open, whether we can assign  any sense beyond the methodological (``toy'') exploration to these model values.

In any case,  Weyl universes, and in particular case 2 equilibrium cosmologies, lead to a striking correspondence with empirical knowledge of different origin and on quite different levels  (supernovae, quasar magnitudes, quasar frequencies, and the  unexpected Pioneer frequency shift). Even if the case 2 Ansatz for self binding energy may not yet be a finally realistic one and  of predominantly   methodological value, we have to keep in mind that  the structural (geometrical) result of Weyl universes seem to  be typical for {\em any} self binding Ansatz which leads to equilibrium  geometries (cf. thm. 1). The empirical  properties of Weyl universes, which have yet only been partially explored, are of much broader import than the specific assumptions of case 2. This is   a sufficient reason for further investigating Weyl geometric equilibrium geometries from various viewpoints.  

Moreover, we should not exclude that scale covariant gravity may become a clue for a better understanding of the coupling of matter and interaction fields to gravity. Drechsler's Higgs mechanism mighS be the first sign from the tip of an iceberg, which is waiting to be discovered.
\\[30mm]

\small
\bibliographystyle{apsr}
\bibliography{a_litfile}
\end{document}